\documentclass[conference]{IEEEtran}
\IEEEoverridecommandlockouts
\usepackage[noadjust]{cite}
\usepackage{amsmath,amssymb,amsfonts}
\usepackage{algorithmic}
\usepackage{graphicx}
\usepackage{textcomp}
\usepackage{xcolor}
\usepackage{url}
\usepackage{hyperref}
\usepackage{adjustbox}
\def\BibTeX{{\rm B\kern-.05em{\sc i\kern-.025em b}\kern-.08em
    T\kern-.1667em\lower.7ex\hbox{E}\kern-.125emX}}
\begin{document}

\title{Static Analysis of Infrastructure as Code: a Survey
\thanks{
This project has received funding from the European Union’s Horizon 2020 programme under grant agreements No 101000162 (PIACERE) and 825480 (SODALITE).\\
\copyright 2022 IEEE. Personal use of this material is permitted. Permission from IEEE must be obtained for all other uses, in any current or future media, including reprinting/republishing this material for advertising or promotional purposes, creating new collective works, for resale or redistribution to servers or lists, or reuse of any copyrighted component of this work in other works.
}}

\author{\IEEEauthorblockN{Michele Chiari}
\IEEEauthorblockA{\textit{DEIB, Politecnico di Milano}\\
Milano, Italy \\
michele.chiari@polimi.it}
\and
\IEEEauthorblockN{Michele De Pascalis}
\IEEEauthorblockA{\textit{DEIB, Politecnico di Milano}\\
Milano, Italy \\
michele.depascalis@mail.polimi.it}
\and
\IEEEauthorblockN{Matteo Pradella}
\IEEEauthorblockA{\textit{DEIB, Politecnico di Milano}\\
\textit{IEIIT, Consiglio Nazionale delle Ricerche}\\
Milano, Italy \\
matteo.pradella@polimi.it}
}

\maketitle
\thispagestyle{plain}
\pagestyle{plain}

\begin{abstract}
The increasing use of Infrastructure as Code (IaC) in DevOps leads to benefits in speed and reliability of deployment operation, but extends to infrastructure challenges typical of software systems.
IaC scripts can contain defects that result in security and reliability issues in the deployed infrastructure: techniques for detecting and preventing them are needed.
We analyze and survey the current state of research in this respect by conducting a literature review on static analysis techniques for IaC.
We describe analysis techniques, defect categories and platforms targeted by tools in the literature.
\end{abstract}


\section{Introduction}

\emph{Infrastructure as Code} (IaC) allows for defining, deploying, managing, and orchestrating computing systems in a fully automated way, based on a textual description of their supporting infrastructure and configuration~\cite{Kief16}.
IaC is part of a larger trend toward the automation of both development and system implementation operations known as \emph{DevOps}~\cite{BassWZ15}.
The integration with widely-used cloud providers offered by IaC platforms makes IaC particularly beneficial for systems leveraging \emph{Cloud Computing}.
Moreover, the benefits of using IaC instead of manual deployments are reduced deployment times, better repeatability, and easier system maintenance, update and management.

On the one hand, representing infrastructure and configuration as code avoids possible errors introduced by human maintainers during deployment operations.
On the other hand, IaC presents features similar to software code~\cite{JiangA15}: IaC code-bases may reach considerable sizes, and are subject to modifications by different maintainers.
Thus, IaC scripts may present similar pitfalls, such as bugs and defects that cause deployment failures or the deployment of defective configurations, that may cause availability, security, performance, or reliability problems.

To prevent and correct such issues, software engineering techniques similar to those developed for software code are increasingly applied to IaC.
These techniques can be divided in dynamic approaches, such as testing and monitoring, which leverage actual deployment executions, and static approaches, which only rely on the analysis of syntactic or semantic features of IaC.
Static analysis can be particularly beneficial for IaC, because it does not require deployment of possibly defective infrastructure, that can be expensive and time-consuming, due to the need of implementing isolated canary environments.

In this paper, we give an overview of recent efforts to develop and apply static analysis techniques to IaC.
We conduct a literature review guided by the following research questions:
\begin{description}
\item[RQ1] What are the tools available for static verification and validation of IaC?
\item[RQ2] What are the techniques used by existing tools?
\item[RQ3] What kinds of properties are checked by existing tools?
\item[RQ4] What IaC languages are targeted by existing tools?
\end{description}
These questions are aimed at gathering an overview of the current state of the art on static techniques for the analysis and detection of anomalies in IaC, to inform researchers on the current trends and highlight research gaps that may lead to future research directions.

As a result, we identify two large categories for IaC static analysis techniques:
\begin{itemize}
\item Those that operate on mostly syntactic features, and rely on code-smell detection or machine learning and data mining.
\item Those that analyze possible behaviors of the infrastructure in different deployment phases, which rely on automated verification techniques such as model checking.
\end{itemize}
We describe the most important works in each category and compare them.
Since a considerable number of papers deals with code smells, we summarize them systematically, to give an overview of defects that may affect IaC.

\smallskip
\emph{Related Work.}
We observe a lack of reviews covering IaC verification literature.
A.~Rahman et al.~\cite{RahmanMW19} do a systematic mapping study of IaC research.
They analyze research on empirical analysis and testing, without focusing on static analysis.
A.~Alnafessah et al.~\cite{AlnafessahGWZCF21} survey work on DevOps quality assurance.
They treat verification in DevOps, but mostly referencing software analysis techniques that are not specific to IaC.

\smallskip
\emph{Paper Structure.}
In Sect.~\ref{sec:background} we briefly introduce IaC;
in Sect.~\ref{sec:methodology} we describe our review methodology;
in Sect.~\ref{sec:code-smells-linters} and \ref{sec:model-checking} we review papers by classifying them in two categories;
in Sect.~\ref{sec:conclusions} we answer the research questions and conclude the paper.

\section{Background: Infrastructure as Code}
\label{sec:background}

IaC is the production of machine-readable textual descriptions of---often cloud---infrastructure that can be automatically executed to deploy and manage computing systems.
IaC platforms provide domain-specific languages for IaC scripts, and offer plugins for integration in cloud service providers and virtualization tools.
Different IaC tools cover different phases of deployment.

Tools such as Ansible\footnote{\url{https://www.ansible.com/}}, Chef\footnote{\url{https://www.chef.io/}}, and Puppet\footnote{\url{https://puppet.com/}} manage the software configuration of computing nodes.
They can deploy software and containers, configure and manage them.
Thus, they often support embedding of system configuration languages such as shell scripts.
Ansible and Chef scripts explicitly describe actions required to reach the target configuration, and can be considered \emph{imperative}, while Puppet describes it in a \emph{declarative} way.
Actions performed by Puppet are \emph{idempotent}, i.e., executing them repeatedly any number of times yields the same results as executing them once.

Platforms such as AWS CloudFormation\footnote{\url{https://aws.amazon.com/cloudformation/}} and Terraform\footnote{\url{https://www.terraform.io/}} focus on provisioning.
They read declarative specifications describing virtual machines and networks connecting them, and synthesize a plan for creating, updating or deleting them.
The plan is then executed to provision the infrastructure.
Deployment tools like Cloudify\footnote{\url{https://cloudify.co/}} can be used to deploy applications on already-provisioned environments.

While each tool targets a different IaC language, the TOSCA~\cite{tosca} language has been introduced to standardize them.
TOSCA allows for a declarative specification of the infrastructure topology as \emph{services templates}, describing concrete or virtual computing nodes, the services they run, and networks connecting them.
Node types can be defined to describe classes of nodes offering certain capabilities, which can be referenced by other nodes.
Deployment plans for TOSCA service templates are created and executed by deployment tools (e.g., Cloudify) and orchestrators (e.g, xOpera\footnote{\url{https://github.com/xlab-si/xopera-opera}}).

For a more complete introduction to IaC, we refer to~\cite{Kief16}.

\section{Methodology}
\label{sec:methodology}

We performed our literature review by searching the main search engines for scientific literature.

We defined the search strings by gathering keywords from the research questions, and by considering possible synonyms and related terms.
This resulted in a query of the form
\(
A \land B
\)
where
\(
A = \text{Infrastructure as Code}
\lor \text{IaC}
\lor \text{TOSCA}
\lor \text{Terraform}
\lor \text{Cloudify}
\lor \text{Puppet}
\lor \text{Chef}
\lor \text{Ansible}
\lor \text{Cloud}
\),
and
\(
B = \text{Verification}
\lor \text{Validation}
\lor \text{Validity}
\lor \text{Static Analysis}
\lor \text{Model Checking}
\lor \text{SMT}
\lor \text{Code Smell}
\lor \text{Linter}
\).

We used the following search engines:
ACM Digital Library, Elsevier Scopus, Clarivate Web of Science, IEEE Xplore Digital Library, and Google Scholar.

After gathering search results, we selected relevant entries by including only those describing IaC static-analysis techniques, and by excluding:
\begin{itemize}
\item papers on verification of cloud infrastructure not specified through IaC;
\item papers employing dynamic techniques such as testing and monitoring;
\item position papers with limited technical content.
\end{itemize}
We sorted the results from search engines by relevance to the query, and stopped after finding 20 consecutive results that would be excluded according to these criteria.

The search on Google Scholar was performed by the second author, and searches on other engines by the first one.
Inclusion of papers selected during these searches was subsequently re-discussed by both authors.

\section{Code Smells and Defect Predictors}
\label{sec:code-smells-linters}

The need to apply software engineering techniques typical of mainstream programming languages to IaC was first highlighted in Y.\ Jiang and B.\ Adams \cite{JiangA15}.
The authors conducted an empirical study on IaC code bases targeting Chef and Puppet, gathered from the OpenStack project repositories.
They found out that IaC code takes up a median proportion of 11\% of code in a repository, which is substantially larger than build files.
Moreover, the distribution of monthly changes to IaC files is similar to that of production code files, although IaC files show a smaller \emph{churn}.%
\footnote{\emph{Churn} is a measure of the size of code changes in a software project. Y.~Jiang and B.~Adams \cite{JiangA15} measure it in lines of code changed in a commit.}
Thus, IaC deserves the same attention as ordinary software in studying, finding and preventing bugs.

In this section, we survey tools and techniques that find potential bugs in IaC scripts by relying mostly on the analysis of syntactic features thereof.
This category includes linters, detectors of code smells and metrics that correlate with defects in IaC scripts.

\subsection{Code Smells}
\label{sec:code-smells}

\emph{Code smells} are code features and patterns that, while not being erroneous \emph{per se}, indicate bad code quality, and strongly correlate with coding mistakes and bugs.
The concept was originally introduced by K.\ Beck and M.\ Fowler \cite{Fowler99}, and has been extensively studied by empirical software engineering researchers, who thoroughly categorize code smells for various programming languages.
We summarize IaC smells found in the literature in Table~\ref{tab:smells}.

In the following, defect detectors are evaluated in terms of \emph{precision}, i.e., the proportion of true positives over all detected issues; and \emph{recall}, i.e., the proportion of detected issues over all actual issues.
The harmonic mean of precision and recall is called \emph{F-score} \cite{TanSKK19}.

The first empirical study on code smells in IaC was carried out by T.\ Sharma et al. \cite{SharmaFS16} on 4,621 GitHub repositories containing Puppet scripts.
The authors manually identified 24 code smells, divided in \emph{implementation smells} (I in Table~\ref{tab:smells}), which affect coding style and formatting issues, and \emph{design smells} (D), which affect the use of abstraction mechanisms offered by Puppet scripts.
They used the \emph{Puppet-Lint} tool augmented with custom rules to detect implementation smells, and developed a tool called \emph{Puppeteer}\footnote{\url{https://github.com/tushartushar/Puppeteer}} to detect design smells, mostly based on code metrics specifically developed for each smell.
They found out that the most common implementation smells are ``improper alignment'', ``improper quote usage'', and ``long statement'', while the most common design smells are ``insufficient modularization'' and ``multifaceted abstraction''.
This study, however, does not investigate the degree of correlation between code smells and actual bugs in IaC scripts.

The results by T.\ Sharma et al. \cite{SharmaFS16} are extended to Chef scripts by J.\ Schwarz et al. \cite{SchwarzSL18}.
They classify code smells in \emph{technology agnostic} (A in Table~\ref{tab:smells}), which are applicable to any IaC platform without altering the detection method substantially, \emph{technology dependent} (E), which can be applied to different platforms after modifying the detection method, and \emph{technology specific} (S), which can be applied to a single IaC language.
They classified code smells from \cite{SharmaFS16} into these categories and provided more general definitions for technology-agnostic ones.
Moreover, they introduced five new technology-agnostic or dependent smells
and two Chef-specific ones.
The authors then augmented the Chef linting tool Foodcritic%
\footnote{Foodcritic custom rules available at \url{https://github.com/swc-rwth/InfrastructureAsCodeSmells}.
We represent this as Foodcritic\textsuperscript{\textdagger} in Table~\ref{tab:smells}.}
to detect the smells they studied, and used it to analyze Chef scripts from the repository of an industrial partner (35 cookbooks) and the official Chef repository (3200 cookbooks).
They found out that the most frequently occurring smells are ``improper alignment'', ``long statement'' and ``misplaced attribute'', which agrees with the findings by T.\ Sharma et al. \cite{SharmaFS16} in Puppet scripts.
Conversely, the ``improper quote usage'' smell is not as frequent in Chef as in Puppet.

A.\ Rahman et al. \cite{RahmanPW19} conduct a study on code smells focused on security issues in Puppet scripts.
They identify 7 security smells by applying \emph{descriptive coding} \cite{Saldana15}, a qualitative analysis technique, to 1,726 Puppet scripts gathered from projects by Mozilla, OpenStack, and Wikimedia.
Each identified smell is associated to a security weakness from the Common Weakness Enumeration (CWE) \cite{cwe}.
Such smells include issues common in contexts other than IaC: ``admin by default'' increases the attack surface by granting excessive access rights to a system, ``hard-coded secret'' exposes user names and passwords that should not be revealed, and ``use of HTTP without TLS'' and ``use of weak cryptography algorithms'' undermine secrecy and integrity of communications by using outdated technologies.
The authors develop SLIC\footnote{\url{https://github.com/akondrahman/IacSec}}, a tool that detects security smells in Puppet scripts by means of rules based on string patterns.
They evaluate SLIC on a dataset of 140 Puppet scripts with manually-identified code smells.
SLIC's overall precision and recall turn out to be very high (both 0.99).
The authors then conduct an empirical study by running SLIC on 15,232 Puppet scripts gathered from GitHub, Mozilla, OpenStack, and Wikimedia repositories.
The most frequent smell is ``hard-coded secret'', present in respectively 21.9\%, 9.9\%, 24.8\%, and 17.0\% of the above mentioned sources.
Other frequent smells are ``suspicious comment'' and ``use of HTTP without TLS''.
To evaluate the actual impact of security smells, they surveyed practitioners by submitting 1,000 bug reports on the detected occurrences, obtaining a 21.2\% response rate.
Practitioners agreed on 69.8\% of the bugs, with a peak agreement of 84.6\% for ``use of weak cryptography algorithms'' and more than 75\% for ``use of HTTP without TLS''.
The smells with least agreement are ``suspicious comment'', slightly above 25\%, and ``default admin'' and ``empty password'', both slightly above 50\%.

A further empirical study is done by F.\ A.\ Bhuiyan and A.\ Rahman \cite{BhuiyanR20} to investigate correlation between the security smells identified in \cite{RahmanPW19}.
The work is carried out on the same dataset and using the same tool as \cite{RahmanPW19}.
The smell pairs that are most frequently co-located are ``hard-coded secret'' and ``suspicious comment'' (which co-locate in up to 11\% of the scripts in one dataset), and ``hard-coded secret'' and ``HTTP without TLS'' (up to 16\%).
The authors also investigate which metrics correlate with security smells:
metrics that correlate most strongly are the \emph{number of lines of code} and \emph{configuration attributes}, and the presence of \emph{hard-coded strings}.

A replication study is done in A.\ Rahman et al. \cite{RahmanRPW21}, where the security smells of \cite{RahmanPW19} are extended to Ansible and Chef scripts.
They identify two more smells, namely ``no integrity check'' and ``missing default case statement'', and develop SLAC, a tool for identifying these smells plus those from~\cite{RahmanPW19} in Ansible and Chef scripts.
SLAC is based on pattern-matching rules, like SLIC, and is evaluated in two rounds on two datasets of Ansible and Chef scripts, where smell instances have been manually identified through \emph{closed coding} \cite{Saldana15}.
The evaluation shows that SLAC has very high precision and recall, both between 0.9 and 1.0 on Ansible and Chef scripts.
An empirical study is also conducted on 4,253 Ansible and 6,070 Chef scripts gathered from OpenStack and GitHub repositories.
The results are similar to those for Puppet scripts in \cite{RahmanPW19}: the most frequent smell is ``hard-coded secret'', which affects from 6.8\% of Chef scripts from GitHub to 22.4\% Ansible scripts from OpenStack, followed by ``suspicious comment'' and ``use of HTTP without TLS''.
A practitioner survey is also conducted by sending bug reports for 500 Ansible and 500 Chef smell occurrences.
The findings are again similar to those for Puppet scripts: ``use of weak cryptography algorithms'' has an agreement well above 75\%, and ``use of HTTP without TLS'' also has a high agreement (100\% for Ansible and a little less than 75\% on Chef).
``Suspicious comment'' still has a very low agreement for Chef scripts (less than 30\%), but it is higher for Ansible (above 75\%).
Additionally, ``unrestricted IP address'' and ``missing default case statement'' have a 100\% agreement on Chef scripts.
We must note, however, that the response rate from practitioners was low (9.4\%), which might undermine the significance of these results.

I.\ Kumara et al. \cite{KumaraVMTHKVK20} develop a tool\footnote{\url{https://github.com/SODALITE-EU/defect-prediction}} for detecting code smells in TOSCA service templates, as part of the SODALITE project\footnote{\url{https://sodalite.eu/}}.
They devise a semantic model of a TOSCA deployment, and create SPARQL \cite{sparql} rules based on it to detect smells.
The smells they detect are mostly security-related.
The tool is evaluated on a case study.

A.\ Rahman et al. \cite{RahmanFPW20} apply descriptive coding to 1,448 defect-related commits from 61 OpenStack repositories containing Puppet scripts and generate a taxonomy of defects.
They identify eight categories:
\begin{itemize}
\item \emph{Conditional}: defects that originate from wrong logic in conditional choices;
\item \emph{Configuration Data}: erroneous---possibly hard-coded---configuration settings;
\item \emph{Dependency}: external artifacts required for the script to work are missing;
\item \emph{Documentation}: code comments and documentation contain outdated or wrong information;
\item \emph{Idempotency}: a script is \emph{idempotent} if executing it repeatedly any number of times yields the same results as executing it one time. If this is not the case, the script violates this property.
\item \emph{Security}: confidentiality, integrity or availability are compromised for the provisioned system;
\item \emph{Service}: defects related to improper provisioning of computing services;
\item \emph{Syntax}: the syntax of the IaC language is violated.
\end{itemize}
Although these are not---strictly speaking---code smells, we include this work in this section because ACID, the tool they develop to detect and categorize defects, is based on similar techniques.
ACID analyzes commit messages and diffs by using a set of pattern-matching rules manually devised by the authors.
An evaluation of ACID on an oracle dataset yields average precision and recall resp.\ of 0.84 and 0.96.
An empirical study on commits from 291 repositories from GitHub, Mozilla, OpenStack and Wikimedia shows that ``configuration data'' is the most frequent defect category, followed by ``syntax'' and ``dependency''.

\begin{table*}
\renewcommand{\arraystretch}{1.3}
\centering
\caption{Code Smells in IaC}
\label{tab:smells}
\scriptsize
\begin{tabular}{p{3.2cm} p{7.6cm} p{.6cm} p{1.7cm} p{1cm} p{1.4cm}}
  \hline
  Smell Name(s) & Description & Cat. & Platforms & Ref. & Tools \\
  \hline
  Admin by Default & Access to a resource is obtained through a user with excessive privileges. E.g., a database is accessed by the admin user. & -- & Chef, Puppet, TOSCA & \cite{RahmanPW19,RahmanRPW21,KumaraVMTHKVK20} & SLIC, SLAC, SODALITE \\
  Avoid Comments & The script contains comments. & I, A & Chef & \cite{SchwarzSL18} & Foodcritic\textsuperscript{\textdagger} \\
  Broken  Hierarchy & Inheritance is not used within the same module. E.g., a resource inherits from one in a different namespace. & D & Puppet & \cite{SharmaFS16} & Puppeteer \\
  Complex Expression & The script contains a long and convoluted expression. & I & Puppet & \cite{SharmaFS16} & Puppeteer \\
  Deficient Encapsulation & A class or module has too many global variables referenced by other classes. & D & Puppet & \cite{SharmaFS16} & Puppeteer \\
  Dense Structure & The overall infrastructure has a dense dependency graph. & D & Puppet & \cite{SharmaFS16} & Puppeteer \\
  Deprecated statement usage & The script uses statements deprecated by the platform maintainers. & I & Puppet & \cite{SharmaFS16} & Puppeteer \\
  Duplicate Block & The script contains a number of consecutive duplicate lines higher than a threshold. & I, A & Chef, Puppet & \cite{SharmaFS16,SchwarzSL18} & Puppeteer, Foodcritic\textsuperscript{\textdagger} \\
  Duplicate Entry & Duplicate hard-coded parameters or property values. & I & Puppet & \cite{SharmaFS16} & Puppeteer \\
  Empty Default & A Chef project lacks a \texttt{default.rb} file or it is empty. & D, S & Chef & \cite{SchwarzSL18} & Foodcritic\textsuperscript{\textdagger} \\
  Empty Password & The empty string is used as a password. & -- & Ansible, Puppet, TOSCA & \cite{RahmanPW19,RahmanRPW21,KumaraVMTHKVK20} & SLIC, SLAC, SODALITE \\
  Hard-coded Secret & Sensitive data (e.g., user names, passwords, SSH keys, etc) are hard-coded into the script. & -- & Ansible, Chef, Puppet, TOSCA & \cite{RahmanPW19,RahmanRPW21,KumaraVMTHKVK20} & SLIC, SLAC, SODALITE \\
  Hyphens & The Chef style guide discourages hyphens in cookbook names. & I, S & Chef & \cite{SchwarzSL18} & Foodcritic\textsuperscript{\textdagger} \\
  Imperative Abstraction & Imperative statements are used in a declarative language. & D & Puppet & \cite{SharmaFS16} & Puppeteer \\
  Improper Alignment & Code indentation is inconsistent, or contains tabs. & I, A & Chef, Puppet & \cite{SharmaFS16,SchwarzSL18} & Puppeteer, Foodcritic\textsuperscript{\textdagger} \\
  Improper Quote Usage & Single and double quotes are used when they should not, or are not used when they should. & I, E & Chef, Puppet & \cite{SharmaFS16,SchwarzSL18} & Puppeteer, Foodcritic\textsuperscript{\textdagger} \\
  Include Consistency & The project contains transitive dependencies between similar modules. E.g., a module A references a module B, and they both reference two modules offering similar features. & D, E & Chef & \cite{SchwarzSL18} & Foodcritic\textsuperscript{\textdagger} \\
  Incomplete Conditional & An \texttt{if} statement lacks an \texttt{else} clause. & I & Puppet & \cite{SharmaFS16} & Puppeteer \\
  Incomplete tasks, Suspicious comments & The code contains ``TODO'' or ``FIXME'' comments. & I & Ansible, Chef, Puppet, TOSCA & \cite{SharmaFS16,RahmanPW19,RahmanRPW21,KumaraVMTHKVK20} & Puppeteer, SLIC, SLAC, SODALITE \\
  Inconsistent naming convention & Naming conventions used in the script deviate from the conventional ones. & I & Puppet, TOSCA & \cite{SharmaFS16,KumaraVMTHKVK20} & Puppeteer, SODALITE \\
  Insufficient Key Size & A cryptography key is smaller than a threshold size. & -- & TOSCA & \cite{KumaraVMTHKVK20} & SODALITE \\
  Insufficient modularization & The size of a class or module is excessive (above a certain threshold). & D, E & Chef, Puppet & \cite{SharmaFS16,SchwarzSL18} & Puppeteer, Foodcritic\textsuperscript{\textdagger} \\
  Invalid IP address binding, Unrestricted IP address & A resource is assigned the IP address \texttt{0.0.0.0}. & -- & Ansible, Chef, Puppet, TOSCA & \cite{RahmanPW19,RahmanRPW21,KumaraVMTHKVK20} & SLIC, SLAC, SODALITE \\
  Invalid Port Ranges & TCP port numbers are not between 0 and 65535. & -- & TOSCA & \cite{KumaraVMTHKVK20} & SODALITE \\
  Invalid Property Value & Property or attribute has forbidden value. E.g., malformed file mode mask. & I & Puppet & \cite{SharmaFS16} & Puppeteer \\
  Law of Demeter & The project has transitive dependencies. E.g., a module A references a module B, and they both reference a module C. & D, E & Chef & \cite{SchwarzSL18} & Foodcritic\textsuperscript{\textdagger} \\
    Long Resource & A resource definition spans an excessive number of lines. & D, A & Chef & \cite{SchwarzSL18} & Foodcritic\textsuperscript{\textdagger} \\
  Long Statement & The script contains long lines (that do not fit in a screen). & I, A & Chef, Puppet & \cite{SharmaFS16,SchwarzSL18} & Puppeteer, Foodcritic\textsuperscript{\textdagger} \\
  Misplaced Attribute & Attributes or properties are sorted differently from the conventional order. & I, A & Chef, Puppet & \cite{SharmaFS16,SchwarzSL18} & Puppeteer, Foodcritic\textsuperscript{\textdagger} \\
  Missing Abstraction & Resources are not encapsulated in appropriate abstractions. & D & Puppet & \cite{SharmaFS16} & Puppeteer \\
  Missing Default Case & Default case missing in a \texttt{switch}, \texttt{case} or \texttt{selector} statement. & I & Chef, Puppet & \cite{SharmaFS16,RahmanRPW21} & Puppeteer, SLAC \\
  Multifaceted Abstraction & An abstraction violates the \emph{single responsibility principle} \cite{Martin03}. E.g., a resource declaration corresponds to more than one physical resource, or elements declared in a module are not cohesive. & D, A & Chef, Puppet & \cite{SharmaFS16,SchwarzSL18} & Puppeteer, Foodcritic\textsuperscript{\textdagger} \\
  No Integrity Check & A file is downloaded without checking it for integrity with, e.g., checksums or GPG signatures. & -- & Ansible, Chef & \cite{RahmanRPW21} & SLAC \\
  Too many Attributes & A resource has to many attributes (above a threshold). & D, A & Chef & \cite{SchwarzSL18} & Foodcritic\textsuperscript{\textdagger} \\
  Unguarded Variable & A variable is not enclosed in braces when used in a string. & I, A & Chef, Puppet & \cite{SharmaFS16,SchwarzSL18} & Puppeteer, Foodcritic\textsuperscript{\textdagger} \\
  Unstructured Module & A module is not structured in the conventional way. & D, E & Chef, Puppet & \cite{SharmaFS16,SchwarzSL18} & Puppeteer, Foodcritic\textsuperscript{\textdagger} \\
  Unnecessary Abstraction & The script contains an empty module or class. & D & Puppet & \cite{SharmaFS16} & Puppeteer \\
  Use of HTTP without TLS, Insecure Communication & Transport Layer Security is not used by default (HTTP instead of HTTPS). & -- & Ansible, Chef, Puppet, TOSCA & \cite{RahmanPW19,RahmanRPW21,KumaraVMTHKVK20} & SLIC, SLAC, SODALITE \\
  Use of weak cryptography algorithms & Deprecated algorithms are used for encryption (e.g., MD4, MD5, SHA1). & -- & Chef, Puppet, TOSCA & \cite{RahmanPW19,RahmanRPW21,KumaraVMTHKVK20} & SLIC, SLAC, SODALITE \\
  Weakened Modularity & A module has a higher proportion of inter-module (i.e., external) references than intra-module (i.e., internal). & D, E & Chef, Puppet & \cite{SharmaFS16,SchwarzSL18} & Puppeteer, Foodcritic\textsuperscript{\textdagger} \\
  \hline
\end{tabular}
\end{table*}

\subsection{Methods based on Data-Mining}

While IaC smells are devised manually by authors of works described in Sect.~\ref{sec:code-smells}, in this section we analyze works where script features indicating potential defects are obtained through systematic qualitative analyses or data and code mining techniques, which are then utilized to create detection tools.

A.\ Rahman and L.\ Williams \cite{RahmanW18} collect 2,259 Puppet scripts from repositories by Mozilla, OpenStack and Wikimedia Commons, and apply qualitative analysis to their commits to identify the ones related to IaC script defects.
Then, features of defective scripts are mined through the \emph{Bag of Words} (BOW) \cite{Harris54} and \emph{Term Frequency-Inverse Document Frequency} (TF-IDF) \cite{ManningRS80} techniques, and the most relevant ones are selected through \emph{Principal Component Analysis} (PCA) \cite{TanSKK19}.
Categories are derived from such text features through a qualitative analysis by \emph{Strauss-Corbin Grounded Theory} (SGT) \cite{CorbinS15}.
They identify three categories: \emph{filesystem operations}, \emph{infrastructure provisioning}, and \emph{managing user accounts}.
Predictors for these categories are built by training \emph{Random Forest} \cite{TanSKK19} statistical learners, which are evaluated by applying \emph{10-fold cross-validation} \cite{TanSKK19} on the same dataset.
The resulting median F-scores range from 0.70 to 0.74.

More fine-grained categories are obtained by A.\ Rahman and L.\ Williams in \cite{RahmanW19}, where they apply constructivist grounded theory to defect-related commit messages and diffs to identify source-code properties of defective scripts.
The dataset consists of 2,439 Puppet scripts from Mirantis, Mozilla, OpenStack and Wikimedia repositories.
Some of the properties they find have also been identified as code smells (cf.\ Sect.~\ref{sec:code-smells}).
The properties that correlate the most with defects are the number of lines of code and hard-coded strings.
The authors then build predictors using five different statistical learners and evaluate them using 10-fold cross-validation, obtaining F-scores between 0.67 and 0.70 for the best techniques.
Predictors are also compared with implementation smells by T.\ Sharma et al. \cite{SharmaFS16}, finding out that predictors are better in precision but worse in recall.

N.\ Borovits et al. \cite{BorovitsKKPNPTH20} developed DeepIaC, a tool that uses deep learning to detect anti-patterns in Ansible scripts.
DeepIaC classifies scripts based on whether they contain anti-patterns by means of Convolutional Neural Networks \cite{GoodfellowBC16} trained on a dataset of scripts with artificially-inserted bugs.
An empirical evaluation on 18,286 scripts taken from 38 GitHub repositories shows that DeepIaC detects such artificial bugs with an accuracy ranging from 0.79 to 0.92.

Code and process metrics are evaluated by S.\ Dalla Palma et al. \cite{DallaPalmaDPT21} as features for machine-learning-based detection of defects in IaC scripts.
The authors mine GitHub repositories and gather a dataset of 104 repositories containing defective Ansible scripts \cite{DallaPalma20}.
Then, they automatically extract from them 108 features taken from previous work on process metrics for general software \cite{RahmanD13,PalmaNPT20} and code metrics specific to IaC \cite{RahmanW19}.
These features are used to train predictors consisting of a feature selection, a data balancing, a data normalization, and a classification phase, using different combinations of techniques for each phase.
They find out that Random Forest is by far the best predictor, scoring first in terms of accuracy and recall on 98 repositories out of 104.
Moreover, IaC-specific metrics greatly outperform other metrics as prediction features.
The resulting predictors are used in the defect-prediction framework of the RADON\footnote{\url{https://radon-h2020.eu/}} project.

S.\ Dalla Palma et al. \cite{PalmaMNT20} notes that in most IaC datasets defective scripts are considerably outnumbered by correct scripts.
This may be an issue for classification approaches based on machine learning, because they need training datasets that contain a sufficiently representative variety of samples in both classes.
Thus, they employ machine-learning techniques for \emph{novelty detection}, which are trained on a dataset of purely correct scripts, and detect defects by finding scripts with features that deviate from the training dataset.
They apply the techniques \emph{OneClassSVN}, \emph{LocalOutlierFactor}, and \emph{IsolationForest} \cite{MarkouS03} to the same dataset of \cite{DallaPalmaDPT21} through 10-fold cross-validation, using only correct scripts for training.
The empirical evaluation yields a mean precision ranging from 0.84 (OneClassSVM) to 0.86 (LocalOutlierFactor and IsolationForest), and a mean recall from 0.70 (LocalOutlierFactor) to 0.77 (OneClassSVM and IsolationForest).
All approaches greatly outperform \emph{RandomForest} \cite{TanSKK19}.

\subsection{Other}

We describe here works that do not fall in one of the two previous categories.

T.\ Dai et al. \cite{DaiKKZ20} developed a tool called SecureCode, which checks shell scripts embedded in or invoked by Ansible scripts.
Although the object of validation are shell scripts and not IaC directly, we include this paper in the review because it highlights the fact that IaC verification tools may need to cope with the fact that IaC often relies on external resources that may present issues too.
SecureCode scans IaC scripts for references to shell scripts, and when it finds shell script templates, it produces concrete scripts for them by instantiating Ansible variables in them.
Then, the scripts are fed to the existing shell-script static analysis tools ShellCheck\footnote{\url{https://www.shellcheck.net/}} and PSScriptAnalyzer\footnote{\url{https://docs.microsoft.com/en-us/powershell/module/psscriptanalyzer/}}.
SecureCode classifies detected issues based on whether they affect security, availability, performance, or reliability.
The authors perform an empirical study on 1,492 scripts from 45 GitHub repositories, on which SecureCode detects 3,535 issues, 116 of which are false positives.
They do not estimate precision and recall.

Sommelier\footnote{\url{https://github.com/di-unipi-socc/Sommelier}}, by A.\ Brogi et al.\ \cite{BrogiTS17}, is a tool aimed at validating relationships between nodes in TOSCA service templates.
The TOSCA standard prescribes that all elements referenced by node relationships must exist in the service template.
Service templates containing undefined references may lead to errors during deployments, if they are not checked.
Thus, the authors give formal definitions of such requirements, and describe conditions in which they are violated, thus allowing for \emph{ad hoc} checks for their validity.

\section{Model checking of IaC}
\label{sec:model-checking}

\textit{Model checking}~\cite{ClarkeHVB18} is a formal-verification approach where an engineered system is modeled in a logical framework in which it is feasible to check that certain desired properties are guaranteed, or whether undesirable situations may occur.
Traditionally, systems are modeled with some kind of graph or transition system, which are natural representations for evolving stateful systems such as electronic devices or imperative programs. Recently, techniques involving SAT (\emph{Boolean Satisfiability}) and SMT (\emph{Satisfiability Modulo Theory}) solvers appeared. They leverage advancements in algorithms for SAT (such as DPLL) to provide tools that often are very fast on practical models, despite the theoretical complexity bounds.

The translation of the verification target into the modeling language can be performed manually by a software architect, or automatically. Manual translation benefits from deep knowledge of the target, but it can be laborious and error-prone. Devising an automatic translator is seldom trivial, and often comes with a cost in generality.

Due to well-known results in Computer Science, there is no automatic procedure capable of singling out all and only the pieces of software, written in a Turing-complete programming language, whose behavior conforms to a given non-trivial property.
Consequently, automatic verification approaches must be more restrictive than it would be desirable in guaranteeing that a program satisfies a requirement.
This does not necessarily apply to IaC, where the concern is rather about the capability of the logical framework to express system properties.
E.g., a model focused on relationships between cloud computing nodes will not be able to predict an internal application malfunction on one of such nodes.

In this section, we describe works in IaC model checking.

K.\ Jayaraman et al.~\cite{jayaraman_automated_2014} developed SecGuru, a tool that analyses firewall ACLs within infrastructures deployed in the Azure cloud.
SecGuru enables inspection of the differences resulting from an ACL update in terms of network packets that are allowed or blocked by the firewall, and to check the behavior of the firewall against a given \textit{policy}.
This is obtained by encoding the action taken by the firewall on packets and the policies into an SMT problem, then solved by the Z3 SMT tool.
The resulting instances correspond to the network packets that are either blocked or allowed.
The tool was evaluated on real and synthetic policies, and is currently active in the Azure cloud, reportedly having a ``measurable positive impact in prohibiting policy misconfigurations''.

A.\ Brogi et al.~\cite{DBLP:conf/esocc/BrogiCS15} propose a tool to verify the validity of TOSCA management plans.
The user enriches each TOSCA node template with a set compatible states.
Then, they characterize management operations and states by specifying the states in which the nodes providing the required capabilities need to be for the execution of the plan to be successful.
The tool checks the validity of a plan by creating a state-representation of the whole infrastructure.
In this representation, a state is a valid combination of the states of individual nodes, and a transition between states is a management operation on a node such that the operation's requirements are satisfied.
The validity of the plan is then equivalent to the existence of an operation-labeled path.

\begin{table*}
\renewcommand{\arraystretch}{1.1}
\centering
\caption{Answers to the RQs}
\label{tab:rq-answers}
\scriptsize
\begin{tabular}{l c l l l}
\hline
Tool name (RQ1) & Ref. & Target Platform (RQ4) & Target Properties (RQ3) & Technique (RQ2) \\
\hline
ACID & \cite{RahmanFPW20} & Puppet & Anti-patterns & Pattern rules \\
Barrel & \cite{DBLP:conf/esocc/BrogiCS15} & TOSCA & Deployment plan correctness & Reachability in Transition Systems \\
DeepIaC & \cite{BorovitsKKPNPTH20} & Ansible & Anti-patterns & Deep Learning \\
Foodcritic with custom rules & \cite{SchwarzSL18} & Chef & Code smells & Pattern rules and code metrics \\
H{\"a}yh{\"a} & \cite{DBLP:conf/tacas/LepillerPSS21} & CloudFormation & Security vulnerabilities & Dataflow graph analysis \\
Puppeteer & \cite{SharmaFS16} & Puppet & Code smells & Pattern rules and code metrics \\
RADON defect prediction framework & \cite{DallaPalmaDPT21} & Ansible & Anti-patterns & Data Mining \\
Rehearsal & \cite{DBLP:conf/pldi/ShambaughWG16} & Puppet & Determinism and idempotency & SMT solving \\
SecGuru & \cite{jayaraman_automated_2014} & Azure ACL & Network policies & SMT solving \\
SecureCode & \cite{DaiKKZ20} & Ansible & Bugs in shell scripts & External tools \\
SLIC & \cite{RahmanPW19} & Puppet & Code smells & Pattern rules \\
SLAC & \cite{RahmanRPW21} & Ansible, Chef & Code smells & Pattern rules \\
SODALITE defect predictor & \cite{KumaraVMTHKVK20} & TOSCA & Code smells & SPARQL rules on OWL2 ontology \\
Sommelier & \cite{BrogiTS17} & TOSCA & Undefined references & \emph{Ad hoc} algorithms \\
-- & \cite{chareonsuk_formal_2016} & TOSCA & LTL on deployed services behavior & Model Checking (SPIN) \\
-- & \cite{PalmaMNT20} & Ansible & Anti-patterns & Data Mining \\
-- & \cite{RahmanW18} & Puppet & Anti-patterns & Data Mining \\
-- & \cite{RahmanW19} & Puppet & Anti-patterns & Data Mining \\
-- & \cite{DBLP:conf/icfem/YoshidaOF15} & TOSCA & Success of orchestrated operations & Interactive Theorem Proving \\
\hline
\end{tabular}
\end{table*}


W.\ Chareonsuk and W.\ Vatanawood~\cite{chareonsuk_formal_2016} present a toolchain to perform formal verification of interacting web services in a TOSCA specification.
The process relies on user-supplied information describing the behavior of the modelled services, written in the \textit{Web Services Business Process Execution Language} (WS-BPEL, or BPEL).
The user provides a BPEL description of services running on the infrastructure, and integrates it in the TOSCA service template using \emph{ad hoc} node and relationship types.
This IaC is then compiled into a PROMELA specification, against which \emph{Linear Temporal Logic} (LTL)~\cite{ClarkeHVB18} formulas can be verified using the SPIN model checker~\cite{Holzmann04}.
As an example, the authors propose checking \textit{safety} properties, expressed as formulas of the form $\Box \neg P$, i.e. ``it is always true that $P$ does not hold'', where $P$ is some undesirable condition.

In \cite{DBLP:conf/pldi/ShambaughWG16}, R.\ Shambaugh et al. developed Rehearsal, a tool to analyze Puppet configurations by compiling them into a formal language describing filesystem operations, and then translating them into SMT specifications. The Z3 SMT solver is used to look for models representing executions that violate the principles of determinism and idempotency. Rehearsal was tested on a small set of 13 Puppet configurations gathered from GitHub and Puppet Forge, and found bugs in 6 of them. Benchmarks on the test set run on a quad-core 3.5 GHz Intel Core i5 with 8GB RAM measured average checking times within 3 seconds for all configurations.

J.\ Lepiller et al.~\cite{DBLP:conf/tacas/LepillerPSS21} identified a class of cloud-related security vulnerabilities, called \textit{intra-update sniping vulnerabilities}. These occur when an infrastructure update operation, despite transitioning between secure states, traverses insecure intermediate states, for instance because components are updated in the wrong order. To detect this vulnerability class in CloudFormation templates, the authors developed H{\"a}yh{\"a},
which models the described infrastructure as a \textit{dataflow graph}.
H{\"a}yh{\"a} was evaluated on a set of open-source CloudFormation templates: while no vulnerability was detected, the tool showed performances acceptable for integration in a deployment workflow, as execution time was within 1 second for all templates.

H.\ Yoshida et al.~\cite{DBLP:conf/icfem/YoshidaOF15} propose a method for manual modeling of TOSCA service templates in the formal specification language CafeOBJ.
This method is useful for proving that orchestration operations can reach the final state of the infrastructure while maintaining a given invariant property in intermediate states.
Proving a property modeled in CafeOBJ is a form of formal verification, but it cannot be regarded as model checking since it requires user interaction.

A number of authors presented techniques for model checking of cloud infrastructure in which the model is constructed manually. H.\ Sahli et al. \cite{DBLP:conf/vecos/SahliBB14} proposed \textit{bigraphical reactive systems} as a suitable logical framework to model cloud infrastructure lifecycles, and exemplified the use of the model checker BigMC to check relevant properties of elasticity and plasticity. K.\ Klai and H.\ Ochi \cite{DBLP:conf/icws/KlaiO16} presented a technique for model checking the interaction of multiple cloud services accessing shared resources concurrently. The cloud services are modeled as \textit{RCoWF} ({\em Resource-Constraint open WorkFlow}) nets and translated into labeled Kripke structures, against which \textit{hybrid LTL} formulae are checked, e.g. to detect deadlocks on a concurrently accessed resource.
We do not describe these works further, because they do not address IaC directly.

\section{Discussion and Conclusions}
\label{sec:conclusions}

We conducted a literature review on IaC static analysis techniques by querying the most important bibliographic search engines.
We summarize the answers to our research questions in Table~\ref{tab:rq-answers}.

Concerning RQ2, we found out that the most used techniques powering the tools are string-pattern rules (5 tools), and machine learning techniques (5 tools).
Model checking is used in 5 tools too, with SMT solvers being most popular.

Regarding RQ3, \emph{anti-patterns}, by which we mean code or process metrics and other features used for training machine learning classifiers, and code smells are the most targeted properties (resp.\ 6 and 5 tools).
Model-checking-based tools often target the runtime behavior of the deployment.

As for RQ4, the most targeted platform is Puppet (6 tools), followed by Ansible (4 tools) and TOSCA (4 tools).

In conclusion, the review shows an increasing attention on quality assurance techniques for IaC.
Code smell and defect detection and prediction techniques have reached considerable advancement, although improvements may be possible in terms of accuracy.
Other future-work lines are investigation of automated remediation strategies for defects, and use of techniques that take the IaC semantics into account, as opposed to currently used pattern-matching and machine-learning.

A direction for future work in model checking tools is to increase the precision of abstractions used for modeling deployments, to enable the verification of more properties.
Better automation of IaC modeling should also be targeted, because several works still employ manual modeling, which is impractical and error-prone.

\bibliographystyle{IEEEtran}
\bibliography{biblio.bib}

\end{document}